\documentclass[aps, superscriptaddress,amsmath,amssymb,twocolumn,amsfonts,floatfix,longbibliography]{revtex4-2}
\usepackage{mathtools}
\usepackage{braket}
\usepackage[dvipsnames]{xcolor}
\usepackage{float}
\usepackage{subfigure}
\usepackage{upgreek}
\usepackage{pgffor}
\usepackage{silence}
\usepackage[colorlinks=true,linktoc=page,linkcolor=blue,citecolor=magenta,urlcolor=Fuchsia]{hyperref}
\usepackage[normalem]{ulem}
\usepackage{sidecap,tikz}
\usepackage{orcidlink}
\usepackage{makecell}
\usepackage{mathrsfs}
\usepackage{comment}
\usepackage{amssymb}
\usepackage{multirow}
\usepackage{makecell}
\usepackage{booktabs}
\usepackage{hhline}

\newcommand{\beq}{\begin{equation}}
\newcommand{\eeq}{\end{equation}}
\newcommand{\bea}{\begin{eqnarray}}
\newcommand{\eea}{\end{eqnarray}}

\newcommand{\eqn}[1]{Eq.~(\ref{#1})}

\newcommand{\sect}[1]{Sec.~\ref{#1}}

\mathchardef\mhyphen="2D

\begin{document}



\title{Quantum-Geometric Spin Magnetization and Spin Currents Across Altermagnetic Symmetry Classes}


\author{Neelanjan Chakraborti
\orcidlink{0009-0004-5710-0739}}
\email{neelanjanc23@iitk.ac.in}
\affiliation{Department of Physics, Indian Institute of Technology, Kanpur 208016, India}

\author{{Snehasish Nandy}}
\thanks{Jointly supervised this work}
\email{snehasish@phy.nits.ac.in}
\affiliation{Department of Physics, National Institute of Technology Silchar, Assam 788010, India}

\author{{Sudeep Kumar Ghosh}\,\orcidlink{0000-0002-3646-0629}}
\thanks{Jointly supervised this work}
\email{skghosh@iitk.ac.in}
\affiliation{Department of Physics, Indian Institute of Technology, Kanpur 208016, India}

\date{\today}

\begin{abstract}
Altermagnets host spin-split band structures despite vanishing equilibrium spin magnetization, making field-induced responses a direct probe of their quantum geometry. Based on symmetry, altermagnets can be classified into M-, S-, A-, and S/A-type according to the presence or absence of inversion and time-reversal symmetries. We develop a unified framework using a generalized quantum geometric tensor that captures both momentum translations and spin rotations of Bloch states. In particular, we introduce the Zeeman counterpart of the conventional spin quantum geometric tensor, whose real and imaginary parts define the Zeeman spin quantum metric and Zeeman spin Berry curvature, respectively. Using this framework, we study linear spin magnetization and spin currents induced by electric and magnetic fields. We show that M- and S-type altermagnets support intrinsic electric-field-induced spin currents and magnetic-field-induced spin magnetization, while A-type altermagnets exhibit both electric and magnetic field induced spin current and spin magnetization. For S/A-type altermagnets, neither inversion nor time-reversal symmetry alone determines the allowed responses. Applying our theory to the S-type $d$-wave altermagnet $\mathrm{FeSb}_2$ and a Rashba-coupled S/A-type $d$-wave altermagnet, we find complete agreement with the symmetry constraints. The predicted responses are sizable and within reach of magnetometry and
spin-transport experiments. Our results establish a symmetry based quantum geometric framework for spin responses in altermagnets and highlight their potential for experimental detection and spintronic applications.

\end{abstract}
\maketitle

\section{Introduction}
Magnetism in solids is conventionally associated with a finite equilibrium spin magnetization arising from the spin polarization of electronic states. Recent advances, however, have uncovered a distinct class of magnetic systems, termed altermagnets, which exhibit vanishing net magnetization despite hosting momentum-dependent spin-split band structures~\cite{Smejkal2022Nature,Smejkal2022PRX,Smejkal2020SciAdv,GonzalezHernandez2021PRL,Lee_2025,Cheong2025,Roig_2024}. This seemingly paradoxical behavior originates from symmetry constraints, such as the combined $C_{n}\mathcal{T}$ symmetry, which enforce cancellation of the spin polarization while allowing momentum-dependent spin splitting~\cite{Smejkal2022PRX,GonzalezHernandez2021PRL}. As a result, altermagnets provide a unique setting in which spin information is encoded not in a net magnetization, but in the momentum-space structure of Bloch states, offering new opportunities to explore spin responses beyond those found in conventional ferromagnets and antiferromagnets~\cite{Smejkal2022Nature,Smejkal2022PRX,Smejkal2020SciAdv,GonzalezHernandez2021PRL,Cheong2025,Roig_2024}. Altermagnets can be classified into four symmetry classes, A-, M-, S-, and S/A-type, according to their magnetic point-group symmetries~\cite{Cheong2025}. All altermagnets break the combined $\mathcal{PT}$ symmetry, while M- and S-type altermagnets preserve inversion symmetry and break time-reversal symmetry, A-type altermagnets preserve an effective time-reversal symmetry and break inversion symmetry, and S/A-type altermagnets break both inversion and time-reversal symmetries~\cite{Cheong2025}.

\begin{figure}[!ht]
\centering
\includegraphics[width=0.99\columnwidth]{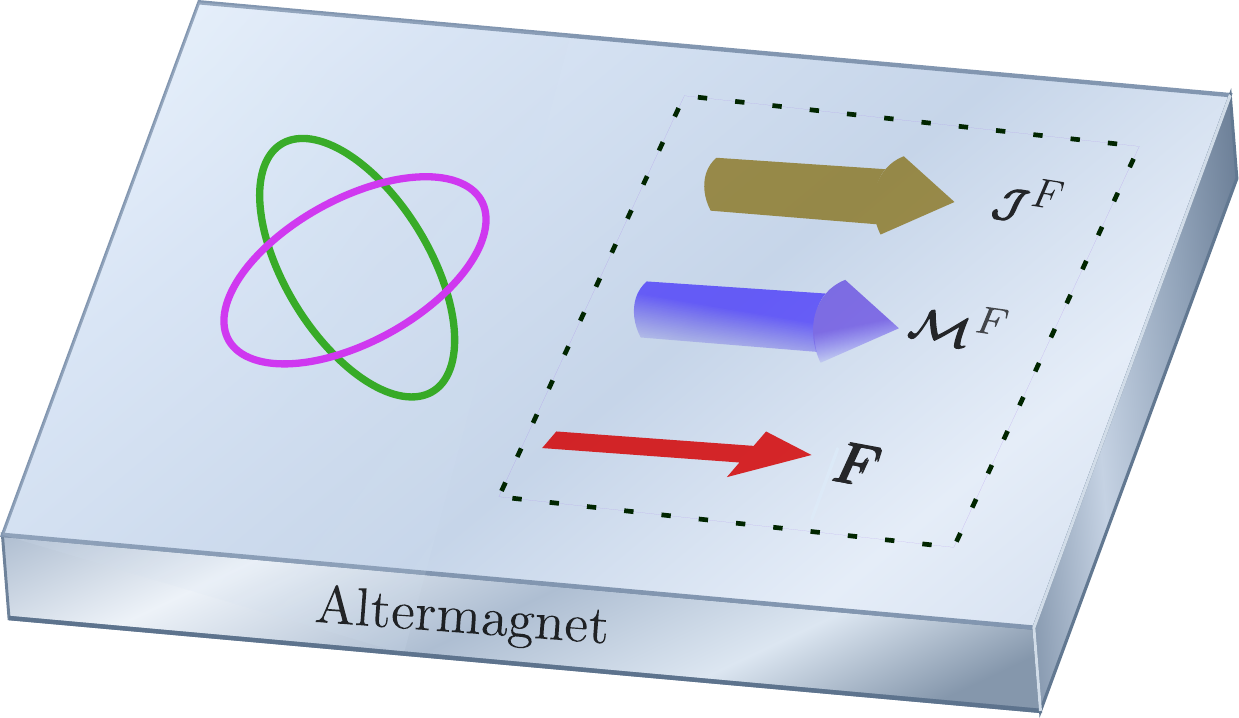}
\caption{\textbf{Field-induced spin magnetization and spin current in an altermagnet:} In the absence of external perturbations ($\mathbf{F}$), an altermagnet exhibits vanishing net spin magnetization despite possessing a spin-split band structure. An applied field ($\mathbf{F}$) induces a finite linear spin magnetization ($\pmb{\mathcal{M}}^F$) and spin current ($\pmb{\mathcal{J}}^F$), whose symmetry-dependent characteristics can distinguish different classes of altermagnets.}
\label{fig:schematic}
\end{figure}

Quantum-geometric effects have emerged as an important framework for understanding unconventional charge and spin responses in magnetic materials~\cite{Jiabin_2025,Liu2024,Ezawa_2026}. In particular, quantum-geometry-induced charge responses in altermagnets have been studied extensively~\cite{Ghorashi_2024, Cheong2025, chakraborti2025}, while the role of quantum geometry in spin magnetization and spin-current responses remains largely unexplored. Recent developments in spin quantum geometry have begun to extend the conventional quantum geometric tensor to spin-dependent observables~\cite{Xiang_2025_PRL,chakraborti2025}, but a systematic understanding of how such geometric quantities govern field-induced spin responses across the different symmetry classes of altermagnets is still lacking. In particular, it remains unclear how momentum translation and spin rotation of Bloch states contribute to electric- and magnetic-field-induced spin responses, and how these contributions are constrained by altermagnetic symmetries. This leads to the central question addressed in this work: {What are the leading-order spin current and spin magnetization responses permitted by symmetry under different external perturbations across distinct classes of altermagnets?}

In this work, we address this question by developing a unified framework for linear spin current and spin magnetization based on the generalized quantum geometry of Bloch states~\cite{Xiang_2025_PRL,chakraborti2025}. We decompose the responses into electric-field-induced and magnetic-field-induced contributions and establish their symmetry constraints across the different altermagnetic classes. Within this framework, momentum translations and spin rotations of Bloch states are treated on equal footing, leading to a generalized quantum-geometric description in terms of the conventional, spin-rotation, and Zeeman spin quantum geometric tensors. In particular, we introduce the Zeeman spin quantum geometric tensor, whose imaginary and real parts define the Zeeman spin Berry curvature and Zeeman spin quantum metric, respectively, and identify these quantities as the geometric origin of magnetic-field-induced spin responses. We find that M- and S-type altermagnets support both intrinsic and extrinsic electric-field-induced linear spin currents, whereas A-type altermagnets permit only an intrinsic electric-field-induced spin current. However, the linear magnetic-field-induced spin-current response vanishes for  M- and S-type classes but remains finite for A-classes. For spin magnetization, M- and S-type altermagnets exhibit an intrinsic magnetic-field-induced linear response governed by the spin-rotation quantum metric, while A-type altermagnets support both an extrinsic electric-field-induced spin magnetization arising from the Zeeman Berry curvature and an intrinsic magnetic-field-induced spin magnetization. For S/A-type altermagnets, where both inversion and time-reversal symmetries are broken, the allowed responses are instead determined by the remaining magnetic point-group symmetries. We verify these symmetry-based predictions using minimal tight-binding models for representative altermagnets, including the S-type $d$-wave altermagnet $\mathrm{FeSb}_2$ and an S/A-type $d$-wave altermagnet with Rashba spin-orbit coupling~\cite{GonzalezHernandez2021PRL,liu2026}. Our results demonstrate that field-induced spin magnetization and spin currents provide a direct route to probing generalized quantum geometry in altermagnets, with potentially large responses in realistic systems.

The remainder of the paper is organized as follows. In \sect{sec:Spin_magnetization}, we introduce the generalized quantum geometry arising from momentum translation and spin rotation of Bloch states, and derive expressions for equilibrium and field-induced spin magnetization and spin current, highlighting the role of the quantum geometry. In \sect{sec:Symmetry_Analysis}, we present a symmetry analysis of the allowed spin magnetization contributions in different altermagnets. In \sect{sec:Results}, we apply our framework to two of the altermagnet candidate materials: S- type $d$-wave altermagnet $\mathrm{FeSb}_2$ and S/A-type $d$-wave altermagnet with Rashba spin-orbit coupling as examples, presenting their spin magnetization and spin current responses. In \sect{sec:Discussions}, we discuss experimental implications, the large magnitude of the predicted spin-magnetization response, and potential spintronic applications. Finally, in \sect{sec:Summary}, we summarize our results and provide an outlook.

\section{Spin Current and Spin Magnetization from Generalized Quantum Geometry of Bloch states}
\label{sec:Spin_magnetization}
In this section, we use the generalized quantum geometric framework and quantum kinetic theory, to derive general expressions for spin magnetization and spin current arising from these quantum geometric quantities in the presence of external electric and magnetic fields.

\subsection{Generalized quantum geometry}
The quantum geometric tensor (QGT), which characterizes the distance between neighboring Bloch states, plays a fundamental role in quantum materials. Its real and imaginary parts correspond to the quantum metric and Berry curvature, respectively. Applying the momentum translation operator $\mathscr{U}_{d\mathbf{k}} = e^{-i d\mathbf{k}\cdot \hat{\mathbf{r}}}$ to a Bloch state $\ket{u^{\zeta}_{m\mathbf{k}}}$ generates $\ket{u^{\zeta}_{m\mathbf{k+d\mathbf{k}}}}$, from which the conventional QGT (CQGT) is obtained as $\mathscr{G}_{mp}^{ab} = r_{mp}^a r_{pm}^b$, where $r^{a}_{mp}(\mathbf{k}) = i \langle u^\zeta_{m\mathbf{k}} | \partial_{k_a} | u^\zeta_{p\mathbf{k}} \rangle$, with $a,b$ denoting spatial components~\cite{Xiao_2010,Jiang_2025,Gao_2014,Du_2021}.

We generalize this construction by additionally applying a spin rotation, $\mathscr{U}_{d\theta} = e^{-i d\boldsymbol{\theta}\cdot\hat{\boldsymbol{\sigma}}/2}$, generated by the spin angular momentum operator~\cite{Xiang_2025_PRL}. The distance between the resulting states is
\begin{align}
    ds^2 &= \left\lVert \ket{ u_{m\mathbf{k+dk}}^{\zeta'}} - \ket {u_{m\mathbf{k}}^\zeta} \right\rVert^2 \nonumber \\
    &= \sum_{p \neq m} \mathscr{G}_{mp}^{ab} \, dk_a \, dk_b
    + \frac{1}{4}\sum_p \mathscr{S}_{pm}^{ab}\,d\theta_a\,d\theta_b \nonumber\\
    &\quad + \frac{1}{2}\sum_{p\neq m}
    \left(\mathscr{Z}_{mp}^{ba}+\mathscr{Z}_{pm}^{ba}\right)
    d\theta_b\,dk_a,
    \label{eq:qgt}
\end{align}
where $\zeta$ and $\zeta'$ denotes the spin index after rotation. Here, $\mathscr{G}_{mp}^{ab}=r_{mp}^a r_{pm}^b$ is the CQGT, while $\mathscr{S}_{mp}^{ab}=\sigma_{mp}^a\sigma_{pm}^b$ and $\mathscr{Z}_{mp}^{ab}=r_{mp}^a\sigma_{pm}^b$ define the spin-rotation QGT (SRQGT) and Zeeman QGT (ZQGT), respectively. Their real and imaginary parts define the corresponding quantum metrics and Berry curvatures: the CQGT gives the conventional quantum metric (CQM) $\mathcal{Q}^{ab}_{mp}$ and Berry curvature (CBC) $\Omega^{ab}_{mp}$; the SRQGT gives the spin-rotation quantum metric (SRQM) $\mathscr{R}^{ab}_{mp}$ and spin-rotation Berry curvature (SRBC) $\Lambda^{ab}_{mp}$~\cite{Jia_2025,chakraborti2026,chakraborti2026_PST}; and the ZQGT gives the Zeeman quantum metric (ZQM) $\mathcal{F}^{ab}_{mp}$ and Zeeman Berry curvature (ZBC) $\Gamma^{ab}_{mp}$~\cite{Xiang_2025_PRL,chakraborti2025_arxiv}.

The CBC and SRBC are purely antisymmetric, while the CQM and SRQM are purely symmetric. In contrast, both the ZBC and ZQM generally contain symmetric and antisymmetric components. In particular, the antisymmetric ZBC satisfies $\sum_p\Gamma_{mp}^{A;ab}=\partial_a\sigma_m^b$, analogous to $\Omega^{ab}_{mp}=\partial_a r_{mp}^b$~\cite{Xiang_2025_PRL}. Under inversion $\mathcal{P}$, the SRBC, CBC, SRQM, and CQM are even, whereas the ZBC and ZQM are odd. Under time reversal $\mathcal{T}$, the SRBC, CBC, and ZQM are odd, while the SRQM, CQM, and ZBC are even.

An important open question concerns the Zeeman counterparts of the geometric quantities governing spin responses. The conventional spin QGT is defined as $T_{nm}^{ab;c} = \frac{v_{nm}^{a;c}v_{mn}^b}{\epsilon_{nm}^2}$, whose real and imaginary parts give the spin quantum metric (SQM), $A_{nm}^{ab;c}$, and spin Berry curvature (SBC), $B_{nm}^{ab;c}$, respectively~\cite{Xiang_2025_spin}. Motivated by this, we introduce the Zeeman spin QGT, $S_{nm}^{ab;c}=v_{nm}^{a;c}\sigma_{mn}^{b}$, whose real and imaginary parts define the Zeeman spin quantum metric (ZSQM), $\mathscr{P}_{nm}^{ab;c}=\mathrm{Re}[S_{nm}^{ab;c}]$, and Zeeman spin Berry curvature (ZSBC), $\mathscr{V}_{nm}^{ab;c}=\mathrm{Im}[S_{nm}^{ab;c}]$, which govern magnetic-field-induced spin transport. Under inversion $\mathcal{P}$, the SBC and SQM are even, while the ZSBC and ZSQM are odd. Under time reversal $\mathcal{T}$, the SQM and ZSQM are odd, whereas the SBC and ZSBC are even. The SBC and SQM are purely antisymmetric and symmetric, respectively, while both the ZSBC and ZSQM generally contain symmetric and antisymmetric components.


\subsection{Quantum Kinetic Theory}

To derive the field-induced spin current and spin magnetization, we start from the quantum Liouville equation~\cite{culcer_2017}, which governs the time evolution of the single-particle density matrix $\rho$:
\begin{equation}
\frac{d\rho}{dt} + \frac{i}{\hbar}[\mathcal{H},\rho] = 0,
\end{equation}
where the total Hamiltonian is given by $\mathcal{H} = \mathcal{H}_0 + \mathcal{H}_F + \mathcal{H}_U$. Here, $\mathcal{H}_0$ denotes the unperturbed band Hamiltonian, $\mathcal{H}_F = - g\mu_B \mathbf{B}(t)\cdot\boldsymbol{\sigma}$ describes the Zeeman coupling to an external AC magnetic field $\mathbf{B}(t) =\mathbf{B}_0 \cos(\omega t)$ and $\mathcal{H}_F = e \mathbf{E}(t)\cdot\mathbf{r}$ represents the perturbation due to an applied AC electric field $\mathbf{E}(t) =\mathbf{E}_0 \cos(\omega t)$. The term $\mathcal{H}_U$ represents the disorder potential. In the weak-disorder limit, within the relaxation-time approximation, the Liouville equation becomes
\begin{equation}\label{eq:DME}
\frac{d\rho}{dt} 
+ \frac{i}{\hbar} \left[\left(\mathcal{H}_0 + \mathcal{H}_F\right), \rho\right] 
+ \frac{1}{\tau} \left[\rho - \rho^{(0)}\right] = 0,
\end{equation}
where $\tau$ is the relaxation time, assumed to be momentum independent for simplicity. We solve Eq.~(\ref{eq:DME}) perturbatively by expanding the density matrix in powers of the external field $(\textbf{F})$ as
$\rho(\mathbf{k}, t)=\rho^{(0)}+\rho^{(1)}(\mathbf{k}, t)+\rho^{(2)}(\mathbf{k}, t)+\cdots$,
where $\rho^{(n)}\propto |\mathbf{F}|^n$ and $\rho^{(0)}$ is the equilibrium density matrix. The first-order correction induced by AC magnetic and electric fields are obtained  as~\cite{chakraborti2026,Xiang_2025_PRL,Bhalla_2022}
\begin{align}
\rho^{(1);B}_{mn} & = \Bigg(-\frac{g \mu_B}{2}\Bigg) 
f_{nm}\,\phi_{mn}\, \sigma^a_{mn} B^a,\nonumber\\
\rho^{(1);E}_{mn} & = e  \Bigg( i \delta_{mn}\partial_af_m \phi_{mm} - f_{mn} r^a_{mn} \phi_{mn}\Bigg) E^a
\label{Eq:rho1}
\end{align}
where $f_{mn}=f_m^0-f_n^0$, $\epsilon_{mn}=\epsilon_m-\epsilon_n$, and $\phi_{mn} = [\frac{i \hbar}{\tau} -\epsilon_{mn} + \hbar \omega]^{-1}$.

\subsection{Spin Magnetization}
Spin magnetization in solids arises from the collective contribution of the spin degrees of freedom of electrons, and can be broadly classified into two categories based on its origin: (i) conventional (equilibrium) spin magnetization that exists in the absence of external perturbations and is given by the macroscopic expectation value of the spin operator over all occupied Bloch states, (ii) field-dependent (non-equilibrium) spin magnetization which emerges in the presence of external fields and originates from both the redistribution of carriers in momentum space and field-induced modifications of the Bloch wave functions.

\subsubsection{Equilibrium Spin Magnetization}
In the absence of an external field, the equilibrium spin magnetization is defined as the expectation value of the spin operator over occupied Bloch states. In systems preserving time-reversal symmetry, it vanishes identically, and a finite value arises only upon symmetry breaking, such as due to intrinsic magnetic order. The equilibrium contribution is given by~\cite{Xiang_2025_spin}
\begin{equation}
\mathcal{M}_a^{(0)} = (g\mu_B)\int_k \sum_n f_n \, \sigma^a_{nn}(\mathbf{k}),
\label{Eq:conventional}
\end{equation}
where $\sigma^a_{nn}(\mathbf{k}) = \langle u^\zeta_{m\mathbf{k}} | \sigma^a | u^\zeta_{m\mathbf{k}} \rangle$ is the spin expectation value of the Bloch state $|u_{n\mathbf{k}}\rangle$, and $f_n = f(\epsilon_n)$ is the Fermi-Dirac distribution. This represents the Brillouin-zone average of the spin density, independent of interband coherence and scattering effects.

\subsubsection{Non-equilibrium Spin Magnetization}
The non-equilibrium spin magnetization in the presence of external fields, such as electric or magnetic fields, can in general be written as 
\begin{equation}
\mathcal{M}_{a} = \beta_{ab} F_b +\beta'_{abc} F_b G_c + ...,
\label{Eq:Linear_Magnetization}
\end{equation}
where $\beta_{ab}$ and $\beta'_{abc}$ are the first and second-order spin magnetization response tensors respectively, induced by the external fields $\mathbf{F}$ and $\mathbf{G}$. In general, both linear and nonlinear response tensors contain intrinsic contributions arising from the geometric properties of Bloch states, and extrinsic contributions, which depend on the relaxation time $\tau$ and capture scattering effects~\cite{Gao2019}.
\paragraph{ Magnetic-field-induced Spin Magnetization:} \mbox{}\\
Under an applied AC magnetic field $\mathbf{B}(t)$, the linear-order spin magnetization introduced in Eq.~(\ref{Eq:Linear_Magnetization}) can be expressed as,
\beq
\mathcal{M}_a^{B} (\omega) = \beta_{ab}^B (\omega) B_b (\omega),
\eeq
where $\beta_{ab}^B (\omega)$ denotes the AC linear spin magnetization response tensor. Using the first-order correction to the density matrix given in Eq.~(\ref{Eq:rho1}), the AC magnetic-field-induced spin magnetization can be written as,
\begin{equation}
\mathcal{M}^{B}_{b} (\omega) = \sum_{nm} \int_k \sigma_{nm}^b \rho^{(1);B}_{mn},
\label{Eq:Magnetizaton}
\end{equation}
where $\rho^{(1);B}_{mn}$ denotes the first-order AC magnetic-field-induced correction to the density matrix. The corresponding frequency-dependent intrinsic spin magnetization response tensor is given by
\begin{align}
\beta^{B;(0)}_{ab} (\omega) & = (g \mu_B) \sum_{nm} \int_k f_n
\Bigg[\frac{\mathscr{R}^{ba}_{nm}}{\epsilon_{mn}} \cos(\omega t) \nonumber \\
& + \frac{2 \hbar \omega}{\epsilon_{nm}^2} \Lambda^{ba}_{nm} \sin(\omega t) \Bigg].
\label{Eq:AC_linear_B}
\end{align}
In the DC limit ($\omega \rightarrow 0$), the linear spin magnetization response tensor in Eq.~\eqref{Eq:AC_linear_B} is given by,
\begin{equation}
\beta^{B;(0)}_{ab} = (g \mu_B) \sum_{nm} \int_k f_n
\frac{\mathscr{R}^{ba}_{nm}}{\epsilon_{mn}} .
\label{Eq:linear_B}
\end{equation}
These expressions show that both the AC and DC magnetic-field-induced spin magnetization responses are entirely intrinsic and independent of the relaxation time. The DC response $\beta^{B;(0)}_{ab}$ originates from a Fermi-sea contribution involving the occupied electronic states. Notably, the magnetic-field-induced spin magnetization response can remain finite even in systems possessing both inversion and time-reversal symmetry. This behavior contrasts with the electric-field-induced spin magnetization, which vanishes in centrosymmetric systems~\cite{Xiang_2025_spin}. This distinction highlights the fundamentally different symmetry requirements of the electric- and magnetic-field-induced spin magnetization responses.

In earlier approaches, magnetization is typically introduced via an explicit Zeeman coupling term in the Hamiltonian, $\mathcal{H}_B = -g\mu_B \,\boldsymbol{\sigma}\cdot\mathbf{B}$, and obtained from $\partial \mathcal{H}/\partial \mathbf{B}$ in the limit $\mathbf{B} \rightarrow 0$~\cite{Yao_2025_prb,CHOI2017513}, where $\mathcal{H} = \mathcal{H}_0 + \mathcal{H}_B$, yielding the equilibrium magnetization discussed earlier. In contrast, here the magnetization originates from a purely geometric mechanism associated with the applied magnetic field and instead emerges intrinsically from the quantum geometry of the Bloch states. \\

\paragraph{Electric-field-induced Spin Magnetization:} \mbox{}\\
The generation of spin magnetization in response to an applied electric field is known as the Edelstein effect~\cite{Xiang_2025_spin,Feng_2025,Kato_2004,EDELSTEIN1990}. To linear order in the electric field $E_b$, the induced spin magnetization, introduced in Eq.~(\ref{Eq:Linear_Magnetization}), can be expressed as
\beq
\mathcal{M}_a^{E} = \beta_{ab}^E E_b,
\eeq
where $\beta_{ab}^E$ is the Edelstein response tensor. This response originates from the field-induced redistribution of carriers in momentum space, which drives the electronic system out of equilibrium and produces an imbalance between states at $\mathbf{k}$ and $-\mathbf{k}$. In systems with spin-split electronic bands, this imbalance can generate a net spin polarization and, consequently, a finite spin magnetization.

Although the Edelstein response can arise from both AC and DC electric fields~\cite{Xiang_2025_spin}, most experimental studies have focused on the DC response~\cite{Sungjae_2020,Senapati2023,Ghiasi_2019}. Using the density matrix given in Eq.~(\ref{Eq:rho1}), the intrinsic and extrinsic contributions to the response tensor under a DC electric field are given by~\cite{Xiang_2025_spin}.
\begin{align}
\beta^{E;(0)}_{ab}
&= e \sum_{nm} \int_k f_n \frac{\mathcal{F}^{ba}_{nm}}{\epsilon_{nm}},
\nonumber\\[4pt]
\beta^{E;(1)}_{ab}
&= -\frac{e\tau}{\hbar}
\sum_{nm} \int_k f_n \, \Gamma^{ba}_{nm}.
\label{Eq:DC_linear_E}
\end{align}
Here, $\mathcal{F}^{ba}_{nm}$ and $\Gamma^{ba}_{nm}$ denote the Zeeman quantum metric and Zeeman Berry curvature, respectively~\cite{Feng_2025}.
These expressions highlight the distinct geometric origins of the Edelstein response. The intrinsic DC response $\beta^{E;(0)}_{ab}$ is governed by the Zeeman quantum metric, whereas the extrinsic DC response $\beta^{E;(1)}_{ab}$ is governed by the Zeeman Berry curvature. Furthermore, the extrinsic contribution is associated with the Fermi surface, since the antisymmetric part of $\Gamma_{nm}^{ba}$ can be related to the momentum derivative of the spin expectation value through $\sum_p \Gamma_{mp}^{A;ab} = \partial_a \sigma^b_m$, as discussed above. By contrast, the intrinsic contribution is a Fermi-sea response, involving contributions from the occupied electronic states.

It is important to note that apart from QGT-induced contribution, a charge current could generate transverse spin polarization via the relation~\cite{EDELSTEIN1990},
\begin{align}
\langle \boldsymbol{\sigma} \rangle = \left(\frac{2 \alpha p_F e\tau}{\hbar^2 \epsilon_F}\right)[\mathbf{J} \times \mathbf{\hat{e}}],
\label{eq:Edelstein}
\end{align}
where $\alpha$, $p_F$, $\epsilon_F$, and $\tau$ denote the Rashba coupling, Fermi momentum, Fermi energy, and scattering time, respectively, and $\mathbf{\hat{e}}$ is normal to the plane. The contribution $\langle \boldsymbol{\sigma} \rangle$ given in Eq.~\eqref{eq:Edelstein} depends explicitly on the Rashba SOC strength $\alpha$. In contrast, the contribution given in Eq.~\eqref{Eq:DC_linear_E} does not explicitly depend on $\alpha$ and contains both intrinsic and extrinsic contributions. Their distinct dependence on the Rashba SOC strength and disorder strength may therefore provide a means of distinguishing the two contributions. In addition to the purely electric-field-induced response, the linear-order spin magnetization can also receive contributions from terms that are jointly linear in the electric and magnetic fields. These mixed responses are governed by the spin-resolved SRQGT and the ZQGT~\cite{Jia_2025}.
Going beyond linear order, second-order spin magnetization induced by an applied electric field has been studied explicitly~\cite{Xiang_2025_spin,Xu_2021,Xiao_2022_new,XU2025100022,Xiao_2023,Sarkar_2025}. Since our discussion is restricted to linear response, we do not consider the second-order electric-field induced contributions here.

\subsection{Spin Current}
Spin current in solids originates from the transport of electronic spin angular momentum and can be broadly classified into two categories according to the driving field: (i) electric-field-induced spin current and (ii) magnetic-field-induced spin current. In both cases, the response receives contributions from two distinct microscopic mechanisms: the redistribution of electron occupations in momentum space and field-induced modifications of the Bloch wave functions. Within the framework of nonlinear response theory, the spin current induced by external fields can be expressed as
\begin{equation}
\mathcal{J}_{a} = \gamma_{ab;c} F_b +\gamma_{abd;c}' F_b G_d + ...,
\label{Eq:Linear_spin_current}
\end{equation}
where $\gamma_{ab;c}$ and $\gamma'_{abd;c}$ are the first- and second-order spin-current response tensors, respectively.

\paragraph{Magnetic-field-induced Spin current:} \mbox{}\\
Under an applied AC magnetic field $\mathbf{B}(t)$, the AC linear spin current introduced in ~\eqn{Eq:Linear_spin_current} can be expressed as
\beq
\mathcal{J}_a^{B} (\omega) = \gamma_{ab;c}^{B} (\omega) B_b (\omega),
\eeq
where $\gamma^{B}_{ab;c} (\omega)$ denotes the AC linear spin current response tensor to an applied AC magnetic field. Using the first-order correction to the density matrix given in Eq.~(\ref{Eq:rho1}), the spin current can be written as
\begin{equation}
\mathcal{J}^{B}_{a} (\omega) = \sum_{nm} \int_k v_{nm}^{a;c} \rho^{(1);B}_{mn},
\label{Eq:Magnetizaton}
\end{equation}
where $v_{nm}^{a;c} = \frac{1}{4} \bra{u_n^\zeta}\{v^a,\sigma^c\}\ket{u_m^\zeta}$ denotes the spin-current matrix element. The only frequency-dependent intrinsic spin current response tensor induced by an AC magnetic field is given by
\begin{align}
\gamma^{B;(0)}_{ab;c} (\omega) & = (g \mu_B)\sum_{nm} \int_k f_n
\Bigg[\frac{\mathscr{P}^{ab;c}_{nm}}{\epsilon_{nm}} \cos(\omega t) \nonumber \\
& + \frac{2 \hbar \omega}{\epsilon_{nm}^2} \mathscr{V}^{ab;c}_{nm} \sin(\omega t) \Bigg].
\label{Eq:linear_B}
\end{align}
The AC spin current response is governed by both the ZSQM and the ZSBC. This intrinsic contribution is independent of the relaxation time $\tau$. Although magnetic-field-induced spin currents have been investigated previously, both theoretically~\cite{Shangguan_2006} and experimentally~\cite{Bozhko_2020}, the specific mechanism identified in our study is completely new.

In the DC limit ($\omega \rightarrow 0$), \eqn{Eq:linear_B} reduces to the linear spin-current response tensor
\begin{equation}
\gamma^{B;(0)}_{ab;c} = (g \mu_B)\sum_{nm} \int_k f_n
\frac{\mathscr{P}^{ab;c}_{nm}}{\epsilon_{nm}}.
\label{Eq:linear_B_spin}
\end{equation}
This expression yields a finite quantum-geometric contribution. However, its physical interpretation remains unclear, as a finite DC response does not necessarily imply a physically realizable steady-state spin-transport current. Interestingly, a finite conduction charge current induced by a DC magnetic field, can also be realized~\cite{Xiang_2025_PRL}. Nevertheless, whether our finite DC spin-current response represents a genuine transport current requires further investigation. Furthermore, this contribution is distinct from the equilibrium background spin currents~\cite{Sonin_2007,Rashba_2003}, as it arises in response to an externally applied magnetic field rather than in the absence of external fields.
\\

\paragraph{Electric-field-induced Spin current:} \mbox{}\\
The generation of spin current in response to an applied electric field is commonly associated with the spin Hall effect~\cite{Xiang_2025_spin}. Under an applied electric field $\mathbf{E}$, the linear-order spin current introduced in Eq.~(\ref{Eq:Linear_spin_current}) can be expressed as
\beq
\mathcal{J}_a^{E} = \gamma_{ab;c}^{E} E_b,
\eeq
where $\gamma^{E}_{ab;c}$ denotes the linear spin current response tensor to an applied electric field. Using the first-order correction to the density matrix given in Eq.~(\ref{Eq:rho1}), the electric-field-induced spin current can be written as
\begin{equation}
\mathcal{J}^{E}_{a} = \sum_{nm} \int_k v_{nm}^{a;c} \rho^{(1);E}_{mn}.
\label{Eq:Magnetizaton}
\end{equation}

In the presence of a DC electric field, the intrinsic and extrinsic contributions to the spin current response tensor are given by~\cite{Xiang_2025_spin}
\begin{align}
\gamma^{E;(1)}_{ab;c}
& =
-2\frac{e\tau}{\hbar}
\sum_{nm}
\int_{\mathbf{k}}
f_n\,\epsilon_{nm}\,
A^{ab;c}_{nm}, \nonumber
\\[4pt]
\gamma^{E;(0)}_{ab;c}
&=
e \sum_{nm}
\int_{\mathbf{k}}
f_n\,
B^{ab;c}_{nm}.
\label{Eq:dc_linear_in_E_spin_current}
\end{align}
The intrinsic contribution, $\gamma^{E;(0)}_{ab;c}$, is governed by the spin Berry curvature, whereas the extrinsic contribution, $\gamma^{E;(1)}_{ab;c}$, is governed by the spin quantum metric and is proportional to the relaxation time $\tau$, reflecting its origin in the nonequilibrium redistribution of carriers. Thus, the intrinsic and extrinsic contributions represent distinct mechanisms underlying the spin current response to an applied electric field.
\\
\begin{table*}[!ht]
\centering
\caption{
\textbf{Summary of symmetry-governed linear spin current and spin magnetization responses in different classes of altermagnets:}
The table summarizes the allowed ($\checkmark$) and forbidden ($\times$) linear spin current and spin magnetization induced by DC electric and AC magnetic fields for different altermagnetic symmetry classes characterized by inversion ($\mathcal{P}$), time-reversal ($\mathcal{T}$), and combined ($\mathcal{PT}$) symmetries. For each allowed response, the corresponding governing quantum-geometric quantity is also indicated.
}
\renewcommand{\arraystretch}{1.55}
\setlength{\tabcolsep}{18pt}

\begin{tabular}{|c|c|cc|cc|}
\hline
\multirow{2}{*}{\makecell{\textbf{Altermagnet}\\ \textbf{class}}}
&
\multirow{2}{*}{\textbf{Symmetry}}
&
\multicolumn{2}{c|}{\textbf{Spin current}}
&
\multicolumn{2}{c|}{\textbf{Spin magnetization}}
\\
\cline{3-6}
&
&
\makecell{\textbf{Electric}\\ \textbf{field}}
&
\makecell{\textbf{Magnetic}\\ \textbf{field}}
&
\makecell{\textbf{Electric}\\ \textbf{field}}
&
\makecell{\textbf{Magnetic}\\ \textbf{field}}
\\
\hline

M-/S-type
&
$\mathcal P$
&
\makecell[c]{
$\checkmark$\\
{\footnotesize SBC}\\
{\footnotesize SQM}
}
&
$\times$
&
$\times$
&
\makecell[c]{
$\checkmark$\\
{\footnotesize SRQM}
}
\\
\hline

A-type
&
$\mathcal T$
&
\makecell[c]{
$\checkmark$\\
{\footnotesize SBC}
}
&
\makecell[c]{
$\checkmark$\\
{\footnotesize ZSBC}
}
&
\makecell[c]{
$\checkmark$\\
{\footnotesize ZBC}
}
&
\makecell[c]{
$\checkmark$\\
{\footnotesize SRQM}
}
\\
\hline

S/A-type
&
$\times$
&
\makecell[c]{
$\checkmark/\times$\\
{\footnotesize SBC, SQM}
}
&
\makecell[c]{
$\checkmark/\times$\\
{\footnotesize ZSBC, ZSQM}
}
&
\makecell[c]{
$\checkmark/\times$\\
{\footnotesize ZBC, ZQM}
}
&
\makecell[c]{
$\checkmark/\times$\\
{\footnotesize SRBC, SRQM}
}
\\
\hline

\end{tabular}
\label{Tab:responses}
\end{table*}

\section{Symmetry Analysis}
\label{sec:Symmetry_Analysis}
There are four classes of altermagnets, namely A-, M-, S-, and S/A-type, distinguished by their magnetic point-group symmetries~\cite{Cheong2025}. All altermagnets break the combined $\mathcal{PT}$ symmetry. M-type and S-type altermagnets break time-reversal symmetry while preserving inversion symmetry. S/A-type altermagnets break both inversion symmetry and time-reversal symmetry. In contrast, A-type altermagnets preserve an effective time-reversal symmetry while breaking inversion symmetry~\cite{Cheong2025}.
For spin-current generation, inversion-symmetric and time-reversal-breaking altermagnets (M- and S-type) allow both intrinsic and extrinsic linear electric-field responses, characterized by nonzero $\gamma^{E;(0)}_{ab;c}$ and $\gamma^{E;(1)}_{ab;c}$. In contrast, inversion-breaking and time-reversal-preserving A-type altermagnets permit only the intrinsic contribution $\gamma^{E;(0)}_{ab;c}$. Notably, the in-phase magnetic-field-induced spin current, governed by the ZSQM, which is odd under both $\mathcal{P}$ and $\mathcal{T}$, vanishes in M-, S-, and A-type altermagnets. The out-of-phase AC component, governed by the ZSBC, which is odd under $\mathcal{P}$ but even under $\mathcal{T}$, vanishes in M- and S-type altermagnets but remains finite in A-type altermagnets. 

\begin{figure*}[!ht]
\centering
\includegraphics[width=\textwidth]{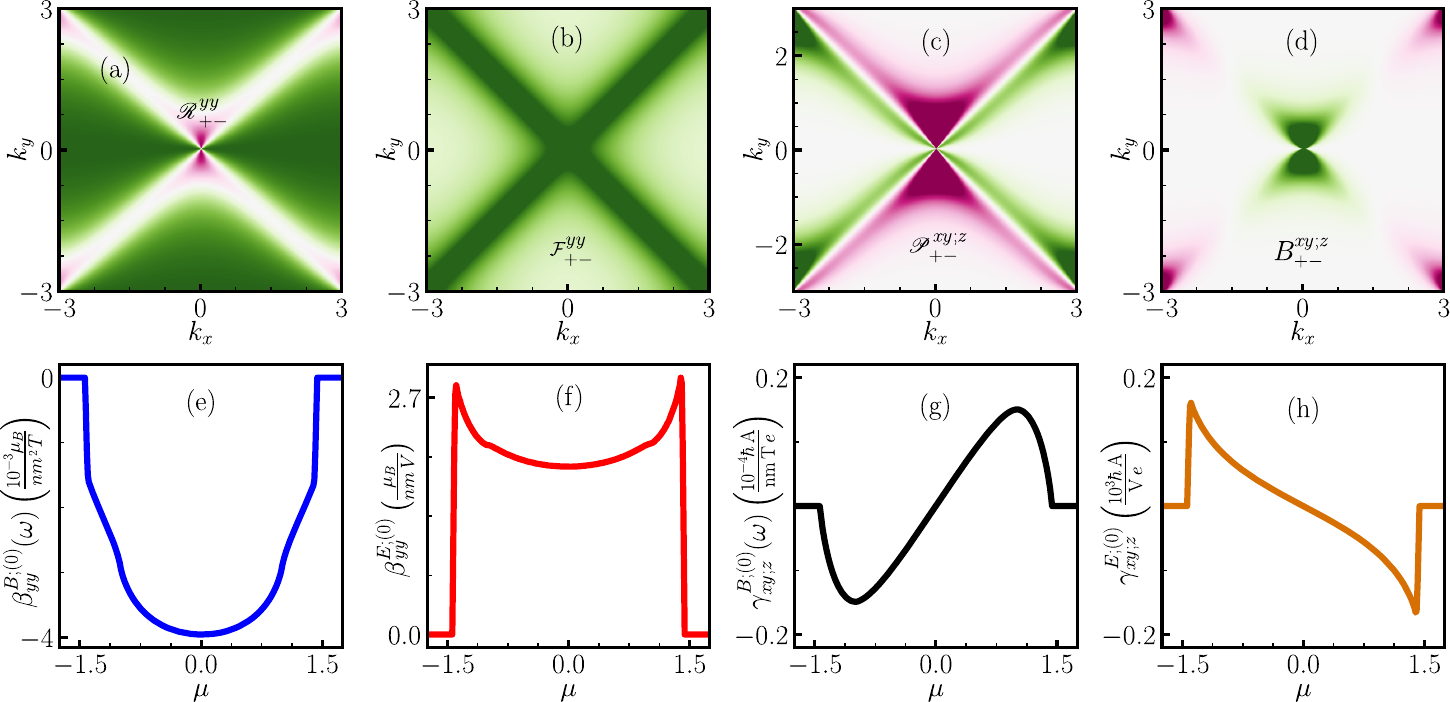}
\caption{\textbf{Spin transport in the $d$-wave altermagnet with Rashba SOC:}
(a) $yy$-component of the spin-rotation quantum metric in the Brillouin zone.
(b) $yy$-component of the Zeeman quantum metric in the Brillouin zone.
(c) $xy;z$-component of the Zeeman spin quantum metric in the Brillouin zone.
(d) $xy;z$-component of the spin Berry curvature in the Brillouin zone.
(e) $yy$-component of the magnetic-field-induced linear spin-magnetization response tensor as a function of chemical potential.
(f) $yy$-component of the electric-field-induced linear spin-magnetization response tensor as a function of chemical potential.
(g) $xy;z$-component of the magnetic-field-induced linear spin-current response tensor as a function of chemical potential.
(h) $xy;z$-component of the electric-field-induced linear spin-current response tensor as a function of chemical potential (in eV).
Panels (a)–(d) show the momentum-space distributions of the geometric quantities responsible for the corresponding responses displayed in panels (e)–(h). Parameters used: $t = 0.35t'$, $\lambda = 0.2t'$, $t_{am} = 0.5t'$ and $t' = 1 \, \text{eV}$.
}
\label{fig:AM_Rashba}
\end{figure*}

For spin magnetization, M- and S-type altermagnets exhibit a linear magnetic-field-induced response described by the intrinsic coefficient $\beta^{B;(0)}_{ab}$. A-type altermagnets, however, support only the electric-field-induced spin magnetization $\beta^{E;(1)}_{ab}$ and the magnetic-field-induced spin magnetization $\beta^{B;(0)}_{ab}$. For S/A-type altermagnets, both inversion symmetry and time-reversal symmetry are broken. Therefore, neither inversion symmetry nor time-reversal symmetry alone can determine whether a given response is allowed or forbidden. The allowed responses are summarized in Table \ref{Tab:responses}.

These results reveal that spin responses can arise at lower order than charge responses in altermagnets. While charge transport generally requires higher-order electric-field responses~\cite{Ghorashi_2024} and linear-order magnetic-field responses~\cite{chakraborti2025}, both spin current and spin magnetization can appear already at linear order.

\section{Results}
\label{sec:Results}
Altermagnets exhibit spin-split band structures despite having vanishing net equilibrium spin magnetization enforced by symmetry~\cite{Roig_2024,Chu_2025,Chen_2025}. To illustrate the symmetry constraints discussed in Sec.~\ref{sec:Symmetry_Analysis}, we consider two representative systems: an S/A-type $d$-wave altermagnet with Rashba SOC and a centrosymmetric S-type $d$-wave altermagnet, $\mathrm{FeSb}_2$.

\subsection{S/A-type $d$-wave altermagnet with Rashba SOC}
We consider a minimal model Hamiltonian for a single orbital on a square lattice with Rashba SOC~\cite{Ghorashi_2024,chakraborti2025}:
\begin{align}
   \mathcal{H}_0 & = -2t [\cos(k_{x}) + \cos(k_{y})] 
   + \lambda [\sin(k_{y}) \sigma_{x} - \sin(k_{x}) \sigma_{y}] \nonumber\\
   & + t_{am} [\cos k_{x} - \cos k_{y}]\sigma_z.
\label{eq:d_wave_rasbha} 
\end{align}
Here, $t$ is the nearest-neighbor hopping amplitude, $\lambda$ is the Rashba spin-orbit coupling strength, $t_{am}$ is the altermagnetic order parameter, and the lattice constant is taken to be unity without loss of generality. The Rashba SOC term breaks inversion symmetry $\hat{\mathcal{P}}$ while preserving time-reversal symmetry $\hat{\mathcal{T}}$, whereas the altermagnetic term preserves inversion symmetry but breaks time-reversal symmetry. Consequently, the Hamiltonian in Eq.~(\ref{eq:d_wave_rasbha}) breaks both $\hat{\mathcal{P}}$ and $\hat{\mathcal{T}}$, and therefore also breaks the combined $\hat{\mathcal{P}}\hat{\mathcal{T}}$ symmetry. The resulting symmetry analysis realizes an S/A-type altermagnetic phase. 

We first consider the spin magnetization. In the altermagnet with Rashba SOC, both electric- and magnetic-field-induced linear spin magnetizations are allowed since no symmetry forbids them. From Eq.~\eqref{Eq:AC_linear_B}, the intrinsic response tensor $\beta^{B;(0)}_{ab} (\omega)$ is entirely governed by the SRQGT. The $\beta^{B;(0)}_{yy} (\omega)$ depends specifically on $\mathscr{R}_{nm}^{yy}$, which exhibits a monopolar momentum-space distribution [Fig.~\ref{fig:AM_Rashba}(a)]. Consequently, $\beta^{B;(0)}_{yy} (\omega)$ maintains the same sign throughout the chemical-potential range, consistent with Fig.~\ref{fig:AM_Rashba}(e). Furthermore, symmetry constraints $\beta^{B;(0)}_{xx} (\omega) = \beta^{B;(0)}_{yy} (\omega)$. From Eq.~\eqref{Eq:DC_linear_E}, the extrinsic linear response tensor $\beta^{E;(1)}_{ab}$ is fully controlled by the ZBC. The band-resolved ZBC satisfies $\Gamma_{nm}^{xy} = -\Gamma_{nm}^{yx}$ and it exhibits a monopolar structure around the center $(k_x,k_y) = (0,0)$. Accordingly, $\beta^{E;(1)}_{xy}$ is finite and symmetry enforces $\beta^{E;(1)}_{xy} = -\beta^{E;(1)}_{yx}$. In contrast, the intrinsic linear response tensor $\beta^{E;(0)}_{ab}$ is governed by the ZQM. Here, $\mathcal{F}_{nm}^{yy}$ remains monopolar over the entire Brillouin zone [Fig.~\ref{fig:AM_Rashba}(b)], leading to a fixed sign of $\beta^{E;(0)}_{yy}$ across all $\mu$ values, consistent with Fig.~\ref{fig:AM_Rashba}(f), while symmetry imposes $\beta^{E;(0)}_{xx} = -\beta^{E;(0)}_{yy}$.

Turning to the spin current, again no symmetry requirement forces the response to vanish, such that both electric- and magnetic-field-induced linear spin currents remain generically finite. From Eq.~\eqref{Eq:linear_B}, the magnetic-field-induced spin-current response tensor $\gamma^{B;(0)}_{xy;z}(\omega)$ is entirely governed by the ZSQM ~[Fig.~\ref{fig:AM_Rashba}(c)], ZSBC part is zero. The resulting response $\gamma^{B;(0)}_{xy;z}(\omega)$ also remains finite, consistent with Fig.~\ref{fig:AM_Rashba}(g).
For the electric-field-driven spin current, both intrinsic and extrinsic mechanisms can contribute; here, we focus exclusively on the intrinsic contribution. As described by Eq.~\eqref{Eq:dc_linear_in_E_spin_current}, the response tensor $\gamma^{E;(0)}_{xy;z}$ is determined by the SBC $B_{nm}^{xy;z}$ [Fig.~\ref{fig:AM_Rashba}(d)]. The predominantly monopolar SBC distribution across the Brillouin zone leads to a finite intrinsic spin-current response. Consequently, $\gamma^{E;(0)}_{xy;z}$ remains nonzero over the entire chemical-potential range, as shown in Fig.~\ref{fig:AM_Rashba}(h).

\subsection{S-type $d$-wave altermagnet FeSb$_2$}
A minimal tight-binding Hamiltonian that captures the essential symmetries of altermagnets and reproduces their characteristic spin-split electronic structure can be written as~\cite{Roig_2024}
\begin{equation}
H = \epsilon_{0,\mathbf{k}}
+ t_{x,\mathbf{k}}\, \tau_x
+ t_{z,\mathbf{k}}\, \tau_z
+ \tau_y\, \boldsymbol{\lambda}_{\mathbf{k}} \cdot \boldsymbol{\sigma}
+ \tau_z\, \mathbf{J} \cdot \boldsymbol{\sigma},
\label{eq:ham}
\end{equation}
where $\tau_i$ and $\sigma_i$ denote the Pauli matrices acting in the orbital (sublattice) and spin spaces, respectively. The term $\epsilon_{0,\mathbf{k}}$ represents the sublattice-independent band dispersion, while $t_{x,\mathbf{k}}$ and $t_{z,\mathbf{k}}$ describe inter- and intra-sublattice hopping amplitudes. The vector ${\lambda}_{\mathbf{k}}$ captures the intrinsic SOC, and $\mathbf{J}$ denotes the N\'eel exchange field associated with the altermagnetic order. The Hamiltonian in Eq.~(\ref{eq:ham}) preserves inversion symmetry while breaking time-reversal symmetry, a prototype S-type altermagnet.

\begin{figure*}
\centering
\includegraphics[width=\textwidth]{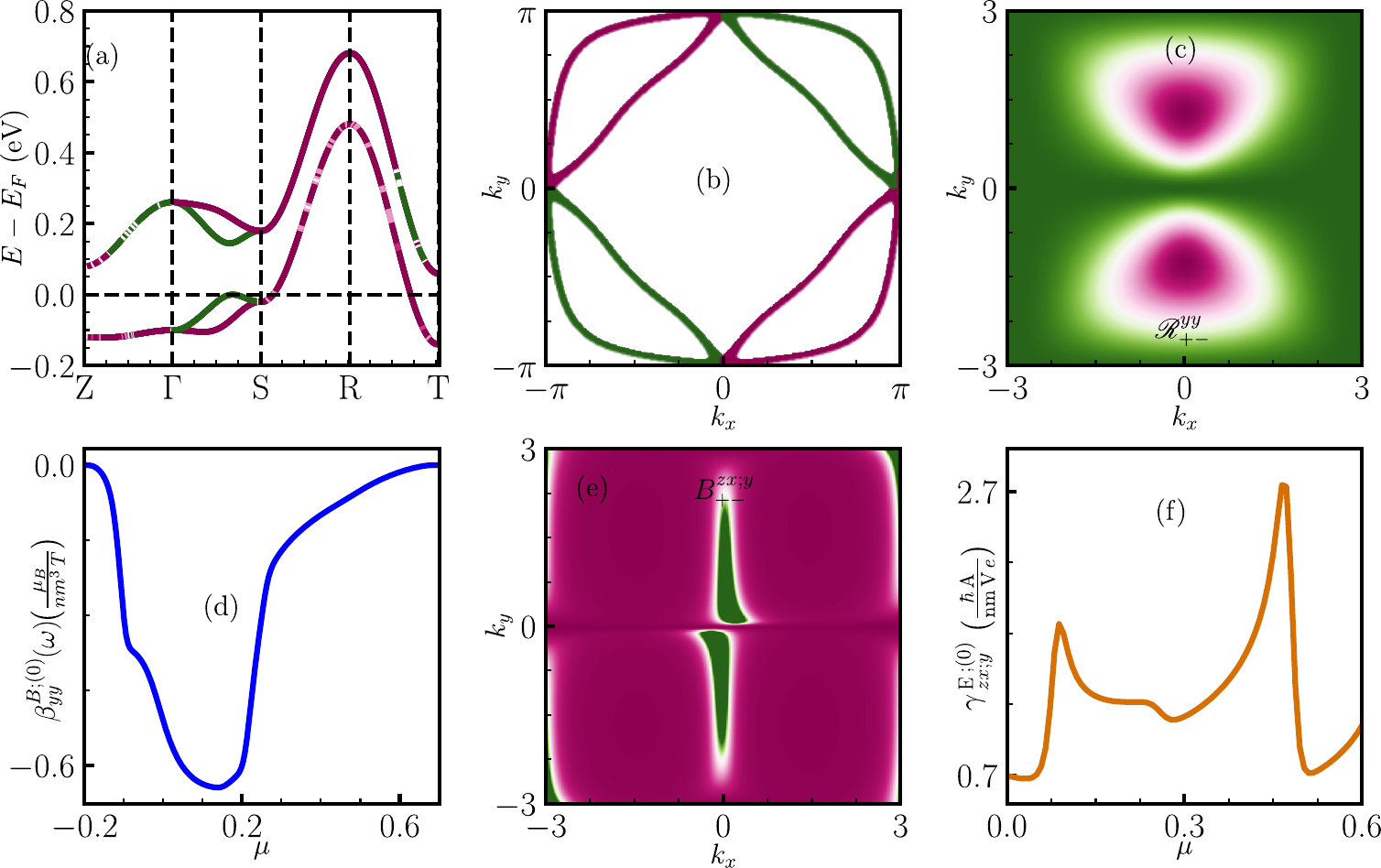}
\caption{\textbf{Spin magnetization in the $d$-wave altermagnet \textbf{FeSb$_2$}:}
(a) Spin-projected band structure along the high-symmetry path $Z$--$\Gamma$--$S$--$R$--$T$, showing pronounced nonrelativistic spin splitting along the $\Gamma$--$S$ and $Z$--$R$ directions.
(b) Spin-projected Fermi surface in the $k_x$--$k_y$ plane at $\mu = 0.16~\mathrm{eV}$, highlighting the characteristic $d$-wave anisotropy of the spin-split bands.
(c) $yy$-component of the spin-rotation quantum metric in the Brillouin zone, revealing the underlying geometric structure.
(d) $yy$-component of the magnetic-field-induced linear spin magnetization response tensor as a function of chemical potential.
(e) $zx;y$-component of the spin Berry curvature in the Brillouin zone.
(f) $zx;y$-component of the electric-field-induced linear spin-current response tensor as a function of chemical potential (in eV).
}
\label{fig:FeSb2}
\end{figure*}

The compound $\mathrm{FeSb}_2$, a possible $d$-wave altermagnet, crystallizes in the centrosymmetric space group $Pnnm$ with lattice constants $a = 5.83\,\text{\AA}$, $b = 6.54\,\text{\AA}$, and $c = 3.18\,\text{\AA}$~\cite{Petrovic_2005,Zhang_2025}. 
Here, the momentum-dependent coefficients entering the Hamiltonian given in Eq.~(\ref{eq:ham}) take the form~\cite{Roig_2024,Sayan_2026_prb}:
\begin{align}
\epsilon_{0,\mathbf{k}} &= 
-\mu 
+ t_{1x}\cos(k_x a) 
+ t_{1y}\cos(k_y b) 
+ t_2 \cos(k_z c) \nonumber \\
&\quad
+ t_3 \cos(k_x a)\cos(k_y b)
+ t_{4x} \cos(k_x a)\cos(k_z c) \nonumber \\
&\quad
+ t_{4y} \cos(k_y b)\cos(k_z c) \nonumber
\\
&\quad
+ t_5 \cos(k_x a)\cos(k_y b)\cos(k_z c), \nonumber \\
t_{x,\mathbf{k}} &\quad= 
t_8 
\cos\!\left(\frac{k_x a}{2}\right)
\cos\!\left(\frac{k_y b}{2}\right)
\cos\!\left(\frac{k_z c}{2}\right), \nonumber \\
t_{z,\mathbf{k}} &\quad= 
\sin(k_x a) \sin(k_y b) [t_6 
+ t_7 \cos(k_z c)], \nonumber\\
\lambda_{x,\mathbf{k}} &\quad= 
\lambda_{x0}
\sin\!\left(\frac{k_x a}{2}\right)
\cos\!\left(\frac{k_y b}{2}\right)
\sin\!\left(\frac{k_z c}{2}\right), \nonumber\\
\lambda_{y,\mathbf{k}} &\quad= 
\lambda_{y0}
\cos\!\left(\frac{k_x a}{2}\right)
\sin\!\left(\frac{k_y b}{2}\right)
\sin\!\left(\frac{k_z c}{2}\right),\nonumber\\
\lambda_{z,\mathbf{k}} &\quad= 
\lambda_{z0}
\cos\!\left(\frac{k_x a}{2}\right)
\cos\!\left(\frac{k_y b}{2}\right)
\cos\!\left(\frac{k_z c}{2}\right),
\end{align}
where 
$t_{1x} = -0.1,\;
t_{1y} = -0.05,\;
t_2 = -0.05,\;
t_3 = 0.06,\;
t_{4x} = 0.1,\;
t_{4y} = 0.05,\;
t_5 = -0.05,\;
t_6 = 0.05,\;
t_7 = -0.1,\;
t_8 = 0.15,\;$
and $J_z = 0.1\;$ (all in eV). We also consider finite in-plane atomic SOC $\lambda_{x0} = \lambda_{y0} = 10~\text{meV}, \lambda_{z0} = 0$.

Figure~\ref{fig:FeSb2}(a) presents the spin-projected band structure of FeSb$_2$ along $Z \rightarrow \Gamma \rightarrow S \rightarrow R \rightarrow T$, where $Z(0,0,\tfrac{1}{2})$, $\Gamma(0,0,0)$, $S(\tfrac{1}{2},\tfrac{1}{2},0)$, $R(0,\tfrac{1}{2},\tfrac{1}{2})$ and $T(\tfrac{1}{2},0,\tfrac{1}{2})$ are the high-symmetry points in the Brillouin zone, revealing pronounced momentum-dependent spin splitting along the $\Gamma$--$S$ high-symmetry direction~\cite{Roig_2024,Sayan_2026_prb}. The corresponding spin-resolved Fermi surface in the $k_x$--$k_y$ plane at $\mu=0.16~\mathrm{eV}$ [Fig.~\ref{fig:FeSb2}(b)] reveals the characteristic $d$-wave anisotropy.

Since the electric-field–driven linear spin magnetization vanishes in centrosymmetric systems, we focus on the magnetic-field–induced response. It is clear from~\eqn{Eq:AC_linear_B} that the intrinsic linear response tensor $\beta^{B;(0)}_{ab} (\omega)$ is governed by the SRQGT. $\beta^{B;(0)}_{yy} (\omega)$ originates from SRQM, here SRBC part is zero. On the $k_z = 0$ plane, the band-resolved SRQM satisfies $\mathscr{R}_{nm}^{xx} = \mathscr{R}_{nm}^{yy}$. The contribution to the SRQM of $n$-th band is defined as $\mathscr{R}_{n}^{aa} = \sum_{m \ne n} \mathscr{R}_{nm}^{aa}$. The $yy$ component of SRQM corresponding to the lowest band, $\mathscr{R}_{1}^{yy}$, exhibits a monopolar distribution in momentum space as shown in Fig.~\ref{fig:FeSb2}(c). Consequently, the corresponding response tensor $\beta^{B;(0)}_{yy} (\omega)$ retains the same sign throughout the entire range of chemical potential $\mu$, in agreement with Fig.~\ref{fig:FeSb2}(d). The symmetry-imposed relation of the SRQM further ensures that $\beta^{B;(0)}_{xx} (\omega) = \beta^{B;(0)}_{yy} (\omega)$. Notably, the spin magnetization is significantly enhanced, exceeding that in conventional antiferromagnets and lying well within experimentally accessible ranges, as discussed in \sect{sec:Discussions}.

The inversion symmetry also restricts the magnetic field induced spin current. Although electric-field-driven spin transport generally contains both intrinsic and extrinsic contributions, we restrict our attention to the intrinsic part. The finite SBC $B_{nm}^{zx;y}$ across a wide region of momentum space [Fig.~\ref{fig:FeSb2}(e)] leads to a nonzero contribution upon Brillouin-zone integration. Consequently, the intrinsic spin-current tensor $\gamma^{E;(0)}_{zx;y}$, defined in Eq.~\eqref{Eq:dc_linear_in_E_spin_current}, remains nonzero, consistent with Fig.~\ref{fig:FeSb2}(f).

\section{Discussion}
\label{sec:Discussions}

Our results establish field-induced spin magnetization and spin current as distinct signatures of different altermagnets, offering clear and experimentally accessible consequences. The symmetry-dependent pattern of allowed and forbidden spin responses provides a distinctive fingerprint of different altermagnetic classes. In particular, the predicted linear spin magnetization can be directly detected through magnetization measurements under weak applied magnetic fields. In SQUID magnetometry, the magnetic moment of the sample is detected through the magnetic flux generated in a superconducting pickup circuit, with the SQUID exploiting Josephson interference to achieve extremely high sensitivity to small magnetic moments~\cite{Clarke1980SQUID,Goree1976SQUID,Perfetti2017Torque,Bartolome2017Torque}. In torque magnetometry, the magnetic moment experiences a torque 
$\pmb{\mathscr{T}}=\mathbf{M}\times\mathbf{B}$~\cite{Jackson1999,Griffiths2017,Kittel2005} in an applied magnetic field,
and the resulting torque is measured as a function of the field orientation, providing information about the magnitude and anisotropy of the field-induced magnetization. These techniques can therefore probe both the magnitude of the response and its dependence on chemical potential, which may be tuned by doping or electrostatic gating. The intrinsic and extrinsic responses can be distinguished experimentally through scaling analysis. While the intrinsic contribution remains insensitive to disorder, the extrinsic contribution scales with the relaxation time $\tau$, allowing their separation by tuning the impurity concentration~\cite{Nandi_2015, Du2019}.


Our calculations predict a spin magnetization density approaching $10^{-2}\,\mu_B/\mathrm{nm}^{3}$ under magnetic fields as small as $10\,\mathrm{mT}$. Likewise, our calculations predict an electric-field-induced spin current approaching $10^{12}\,\hbar A/(e\,\mathrm{m}^{2})$ for electric fields of only $1\,\mathrm{V}/\mu\mathrm{m}$. These responses are enhanced by the interplay of nonrelativistic and relativistic spin splittings, which increases the total spin splitting and amplifies the response, consistent with recent studies~\cite{Yarmohammadi2026,sayan_spin_2026,Sarkar_2025}.

An equally important signature is the symmetry-enforced absence of linear spin-current or spin-magnetization responses in specific altermagnetic classes. This prediction can be tested using current-induced spin-polarization measurements, Kerr rotation, or spin-resolved optical probes~\cite{Soumyanarayanan2016NatPhys}. In such cases, the leading contribution is expected to emerge only at higher order in the applied field, accessible through second-harmonic generation and nonlinear Kerr-effect measurements~\cite{Boyd2008NonlinearOptics,Ma2019Nature}, with characteristic symmetry-constrained tensor anisotropies~\cite{Sodemann2015PRL}.

While our calculations assume an external magnetic field, an equivalent effective magnetic field can be realized through exchange bias in antiferromagnet/ferromagnet heterostructures, where uncompensated interfacial spins generate an internal field of order $B\sim 10^{-3}\!-\!10^{-4}\,\mathrm{T}$~\cite{Wu2022,Jacob_2024}. Extending this approach to altermagnet-based heterostructures provides a practical route to generate and control the predicted response without externally applied magnetic fields.

\section{Summary and outlook}
\label{sec:Summary}
In this work, we established that different symmetry classes of altermagnets exhibit distinct field induced spin magnetization and spin-current responses. By formulating these responses within a generalized quantum geometric tensor framework, we derived symmetry constraints governing the allowed linear responses and identified the Zeeman counterpart of the spin quantum geometric tensor. Its real and imaginary parts define the Zeeman spin quantum metric and Zeeman spin Berry curvature, respectively, providing a unified geometric description of field-induced spin responses.

For electric-field-induced spin currents, M- and S-type altermagnets allow both intrinsic and extrinsic contributions arising from the spin Berry curvature and spin quantum metric, respectively, whereas A-type altermagnets permit only the intrinsic contribution. The linear magnetic-field-induced spin current vanishes in M- and S-type altermagnets but remains finite in A-type altermagnet. For spin magnetization, M- and S-type altermagnets exhibit an intrinsic magnetic-field-induced response governed by the spin-rotation quantum metric, while A-type altermagnets support both an extrinsic electric-field-induced contribution from the Zeeman Berry curvature and an intrinsic magnetic-field-induced contribution. For S/A-type altermagnets, where both inversion and time-reversal symmetries are broken, neither symmetry alone determines the allowed responses.

We verify these symmetry-based predictions using minimal tight-binding models of representative altermagnets, including the S-type $d$-wave altermagnet $\mathrm{FeSb}_2$ and an S/A-type $d$-wave altermagnet with Rashba spin-orbit coupling. The resulting responses are sizable, demonstrating that quantum geometry provides a measurable mechanism for generating spin magnetization and spin currents. Our results therefore establish these responses as experimentally accessible probes of altermagnetic symmetry through magnetometry, transport, and magneto-optical measurements, while providing a framework for exploring quantum-geometric spin phenomena in dynamical, interacting, and topological systems.

\section{Acknowledgments}
N.~C. thanks Dibyendu Samanta for useful discussions and acknowledges the Council of Scientific and Industrial Research (CSIR), Government of India, for providing SRF fellowship. S.~K.~G. and S.~N. acknowledge financial support from Anusandhan National Research Foundation (ANRF) erstwhile Science and Engineering Research Board (SERB), Government of India respectively via the Startup Research Grant: SRG/2023/000934 and the Prime Minister's Early Career Research Grant: ANRF/ECRG/2024/005947/PMS. N.~C. and S.~K.~G. utilized the \textit{Andromeda} server at IIT Kanpur for numerical calculations.

\bibliography{Spin}

\end{document}